\documentclass[pra,twocolumn,showpacs,preprintnumbers,amsmath,amssymb,superscriptaddress]{revtex4-1}

\usepackage{graphicx}
\usepackage{dcolumn}
\usepackage{bm}
\usepackage{epsfig}
\usepackage{amsmath}
\usepackage{amssymb}
\usepackage{color}
\usepackage{bbm}
\usepackage[caption=false]{subfig}
\usepackage{placeins}
\usepackage{soul}

\usepackage[colorlinks=true,citecolor=blue,linkcolor=blue,urlcolor=blue]{hyperref}
\usepackage{cancel}

\newcommand{\ket}[1]{|#1\rangle}

\newcommand{\erw}[1]{\langle#1\rangle}

\newcommand{\abs}[1]{\lvert#1\rvert}

\newcommand{\sz}{\sigma^{z}}
\newcommand{\sx}{\sigma^{x}}

\def\equationautorefname~#1\null{%
	Eq.~#1\null
}
\def\figureautorefname~#1\null{%
	Fig.~#1\null
}
\graphicspath{{./figs/}}

\begin{document}

	\title{The Discrete Truncated Wigner Approximation for Open Quantum Spin System (DTWOQS)}
	
	\title{Realistic simulations of spin squeezing and cooperative coupling effects in large ensembles of interacting two-level systems}   
        
        	\author{Julian Huber}
	\affiliation{Vienna Center for Quantum Science and Technology,
		Atominstitut, TU Wien, 1040 Vienna, Austria}
	\author{Ana Maria Rey}
	\affiliation{JILA, National Institute of Standards and Technology and University of Colorado, 440 UCB, Boulder, Colorado 80309, USA}
	\affiliation{Center for Theory of Quantum Matter, University of Colorado, Boulder, CO, 80309, USA}
	\author{Peter Rabl}
	\affiliation{Vienna Center for Quantum Science and Technology,
		Atominstitut, TU Wien, 1040 Vienna, Austria}

	\date{\today}
	
\begin{abstract}

We describe an efficient numerical method for simulating the dynamics of interacting spin ensembles in the presence of dephasing and decay. The method builds on the discrete truncated Wigner approximation for isolated systems, which combines the mean-field dynamics of a spin ensemble with a Monte Carlo sampling of discrete initial spin values to account for quantum correlations. Here we show how this approach can be generalized for dissipative spin systems by replacing the deterministic  mean-field evolution by a stochastic process, which describes the decay of coherences and populations while preserving the length of each spin. We demonstrate the application of this technique for simulating nonclassical spin-squeezing effects or the dynamics and steady states of cavity QED models with a hundred-thousand interacting two-level systems. This opens up the possibility to perform accurate real-scale simulations of a diverse range of experiments in quantum optics or with solid-state spin ensembles under realistic laboratory conditions.
\end{abstract}

\maketitle
\section{Introduction}
Modeling and understanding the behavior of large ensembles of interacting spins is important for many areas of physics. Apart from more traditional fields, such as magnetism in solids, this includes many recent experiments with cold atoms~\cite{Esteve2008,Appel2009,Riedel2010,SchleierSmith2010,Hosten2016,Cox2016,Pezze2018,Pedrozo2020,Szigeti2020}, trapped ions~\cite{Bollinger2012,Islam2013,Richerme2014,Bollinger2016,Zhang2017,Friis2018,Bruzewicz2019}, Rydberg atoms~\cite{Labuhn2016,Bernien2017,GuardadoSanchez2018,Madjarov2020,Browaeys2020,Scholl2020,Ebadi2020}, polar molecules~\cite{Bohn2017}, magnetic atoms \cite{Lepoutre2019,DePaz2013,Patscheider2020,Burdick2016} or hybrid quantum systems~\cite{Schuster2010,Kubo2010,Amsuss2011,Probst2013,Xiang2013,Kurizki2015,Angerer2018}.  In such settings, an accurate control over a large number of effective spin systems and their coupling to other bosonic degrees of freedom can now be achieved and used for quantum sensing and other quantum technology applications. However, due to the exponential growth of the Hilbert space with increasing number of spins, exact numerical simulations of such systems are typically restricted to only a few tens of spins, which makes a direct theoretical modeling and benchmarking of such experiments impossible. Numerical simulations become even more challenging when realistic decoherence processes are taken into account and the dynamics of the full system density operator must be evaluated.  

In certain situations the exponential scaling in numerical simulations can be avoided and the simulation of moderately large spin systems is still possible. For example, in one dimensional lattices, time-dependent density matrix renormalization group (t-DMRG) techniques can be used to substantially reduce the computational complexity.  This permits the simulation of the coherent~\cite{Daley2004,White2004,Vidal2004} and dissipative dynamics~\cite{Verstrate2004,Orus2008,Mascarenhas2015,Cui2015} of rather large spin chains and with additional effort extensions to two-dimensional lattices are possible~\cite{Kshetrimayum2017,Kilda2020,Keever2020,Yang2020}.  Another important case are systems with a permutational symmetry, for example, a cloud of atoms that couple collectively to a single cavity mode, but also decay individually with the same rate. In this situation the permutation symmetry can be exploited to perform numerical simulations that scale only polynomially with the number of two-level systems~\cite{Chase2008,Xu2013,Kirton2017,Shammah2018,Zhang2018,Shankar2021}. When combined with Monte Carlo wavefunction techniques, the simulation of cavity QED systems with hundreds of atoms~\cite{Kirton2017} or bare ensembles of about $\sim10^5$ two-level systems~\cite{Zhang2018} become possible. However, this latter approach is very restricted and cannot be applied to systems with short-range interactions or, in general, to describe realistic experiments with inhomogeneous frequencies or spatially varying fields. To model such generic  experimental situations it is necessary to identify approximate numerical techniques that take all relevant coherent and incoherent processes accurately into account, but  are still efficient to implement. 
 
In this paper we describe such a general scheme for simulating the dynamics of interacting spin ensembles and cavity QED setups in the presence of dephasing and decay. The method builds on the discrete truncated Wigner approximation (DTWA) introduced in Ref.~\cite{Schachenmayer2015}, where the coherent evolution of interacting spins is approximated by an average over a set of classical trajectories. By taking the exact amount of quantum noise in the initial distribution of those trajectories into account, this technique goes considerably beyond mean-field and provides very accurate predictions for spin-squeezing effects~\cite{Zhu2019,Perlin2020} or for the numerical simulation of quench dynamics~\cite{Czischek2018,Khasseh2020} of systems with hundreds or thousands of spins. However, the DTWA as well as closely related continuous TWA techniques~\cite{Polkovnikov2010} are only applicable for coherent systems. A truncated Wigner method for open quantum spin systems (TWOQS) has recently been introduced in Ref.~\cite{Huber2021} and used to study non-equilibrium phase transitions~\cite{Huber2020}, but this technique is in general only applicable for large collective spins.
Thus, with these existing methods an accurate simulation of many real experiments with dissipation and decoherence is still not possible.

To overcome this problem we present the dissipative discrete truncated Wigner approximation (DDTWA), an extension of the DTWA for open quantum systems, which takes decay and different types of dephasing processes fully into account. This can be achieved by replacing the mean-field dynamics of the classical spin variables by a set of stochastic trajectories. These stochastic equations describe the decay of coherences and populations, but also include the correct amount of added noise. In previous works, such noise processes have been derived based on a semiclassical treatment of the underlying quantum Langevin equations~\cite{Liu2020,Jaeger2021}, which, however, results in  growing or decaying spin fluctuations over time. Here instead we identify appropriate noise processes that preserve the average length of each individual spin. This is the essential ingredient for the accuracy of the DTWA and allows us to simulate also the long-time dynamics and even the steady states of large spin systems in situations where both coherent and dissipative interactions are relevant. Further, the stochastic dynamics of the spins can be readily combined with other  phase space techniques for continuous variable degrees of freedom~\cite{Polkovnikov2010}. Therefore, the method can be immediately adapted for the simulation of ensemble cavity QED settings that include the coupling to lossy photonic modes. At the same time the actual numerical simulations are both straightforward to implement and efficient, such that they can be readily applied for modelling realistic experiments with thousands or even millions of two-level systems. 

The remainder of the paper is structured as follows. In Sec.~\ref{sec:DTWA} we first briefly summarize the DTWA technique for simulating the coherent dynamics of interacting spin ensembles, which we then generalize in Sec.~\ref{sec:stochastic} to take dephasing and decay processes into account. In Sec.~\ref{sec:examples} we illustrate and benchmark the method in terms of a few basic examples, for which exact solutions for comparison still exist. Finally, in Sec.~\ref{sec:largescale} we demonstrate the application of this technique for simulating superradiant decay processes in interacting and inhomogeneous cavity QED systems, for which exact simulation methods are no longer available. A summary of our findings is given in Sec.~\ref{sec:conclusion}.

\section{The discrete truncated Wigner approximation}\label{sec:DTWA}
We are interested in the time evolution of interacting spin ensembles and cavity QED setups with $N\gg1$ effective spin-$1/2$ systems. For concreteness we will first focus on  
pure spin systems described by a Hamiltonian of the form $(\hbar=1)$
\begin{equation}
\mathcal{H}= \frac{1}{2}\sum_{i=1}^N \vec \Omega_i\cdot  \vec \sigma_i + \frac{1}{2} \sum_{i\neq j}^N \vec \sigma_i^{\,T} {\bf J}_{ij} \vec \sigma_j.
\label{eq:spinham}
\end{equation}
Here $\vec \sigma=(\sigma^x,\sigma^y,\sigma^z)^T$, where the $\sigma^k$ are the usual Pauli operators, and $\vec \Omega_i$ and  ${\bf J}_{ij}$ are the local fields and the spin-spin interaction matrix, respectively. Later we will also consider additional couplings of the spin ensemble to a common bosonic mode, as encountered in cavity QED.  Even without the bosonic mode, 
the spins evolving under the action of $\mathcal{H}$ will in general get entangled over time and exact numerical simulations of the full quantum state of the system are only possible for a few tens of spins.  

In Ref.~\cite{Schachenmayer2015} the DTWA was introduced as an approximate numerical method to simulate the coherent dynamics of interacting spin systems.
The basic idea behind this method is to approximate the exact dynamics of the spin ensemble by a set of $N$ classical spin trajectories, $\vec s_i(t)$, which evolve according to the mean-field equations of motion. However, the initial values for these trajectories are randomly drawn from a probability distribution that accounts for the correct quantum mechanical uncertainties of the initial spin state. This leads to a significant improvement compared to mean-field theory.
%
 
\begin{figure}[t]
	\centering
	\includegraphics[width=1\columnwidth]{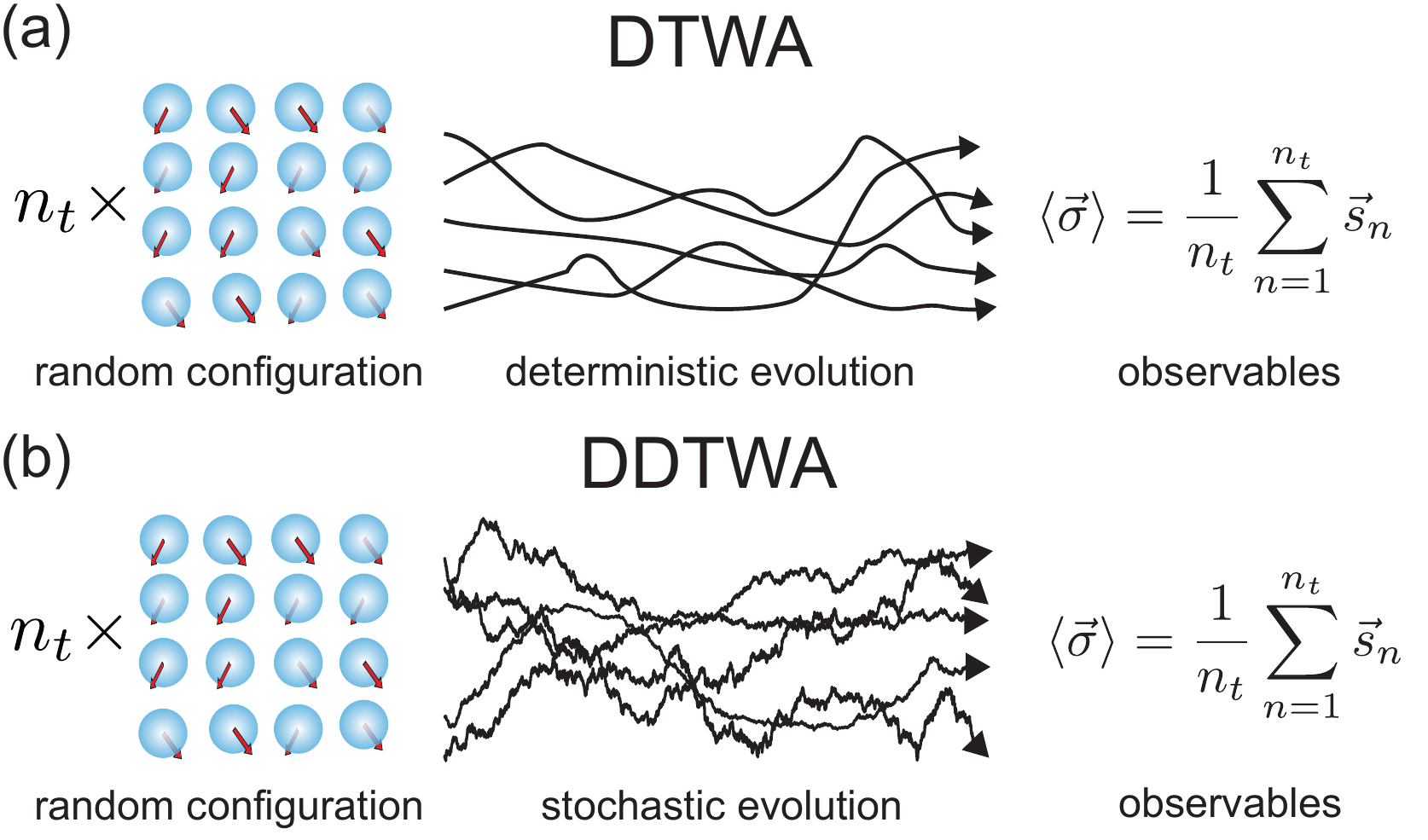}
	\caption{Illustration of (a) the DTWA algorithm~\cite{Schachenmayer2015} for coherent spin systems and (b) the DDTWA algorithm introduced in this work for open quantum spin systems. See text for more details.}
	\label{fig:DDTWAscheme}
\end{figure}

 The actual numerical simulation is performed by implementing the following steps [see Fig.~\ref{fig:DDTWAscheme}(a)]:
\begin{enumerate}
	\item Draw a set of $N$ classical spin variables $\vec s_i=(s_i^{x}, s_i^{y}, s_i^{z})$ according to the discrete Wigner distribution $W_{D}(\{ \vec s_i \})$~\cite{Wooters1987}. 
	For example, for a single spin pointing down, $|\downarrow\,\rangle$, we have~\cite{Pinero2017} 
	\begin{equation}
	\begin{split}
	W_{D}(\vec s_i)=&\frac{1}{4} \delta(s^z_i+1) \left[\delta(s_i^x+1)+\delta(s_i^x-1)  \right] \\  
	& \times \left[\delta(s^y_i+1)+\delta(s^y_i-1)\right].
	\end{split}
	\end{equation}
	This means that the initial spin vectors are randomly drawn from one of the four spin configurations
	\begin{equation}\label{eq:Configurations}
	(s_i^x,s_i^y,s_i^z)=(\pm 1,\pm 1,-1),
	\end{equation}
	which occur with the same probability of $1/4$. All other states on the Bloch sphere can be sampled using the same configurations, followed by an appropriate rotation~\cite{Pinero2017}. 
	
	
	\item Evolve the classical spins according to the mean-field equations of motion, which for the Hamiltonian given in Eq.~\eqref{eq:spinham} read
	\begin{equation}
	\frac{d\vec s_i}{dt}= \vec \Omega_{\rm eff}^i \times \vec s_i, \qquad \vec \Omega_{\rm eff}^i= \vec \Omega_i +  2 \sum_{j=1}^N {\bf J}_{ij} \vec s_j.
	\label{eq:MF}
	\end{equation}
	
	\item Repeat steps 1 and 2  for $n_t \gg1$ times. Expectation values of (symmetrically-ordered) spin observables are then calculated from the average over all trajectories as
	\begin{equation}
	\erw{\sigma_i^k}(t)  \simeq  \frac{1}{n_t} \sum_{n=1}^{n_t} s_{i,n}^k(t),
	\end{equation}
	and
	\begin{equation}
	\erw{\{\sigma_i^k\sigma_j^\ell\}_{\rm sym}}(t)  \simeq  \frac{1}{n_t} \sum_{n=1}^{n_t} s_{i,n}^k(t) s_{j,n}^\ell(t),
	\end{equation}
        where $\{\cdot \}_{\rm sym}$ denotes the symmetrically-ordered operator product and the $ \vec s_{i,n}(t)$ denote the classical spin vectors along the $n$-th trajectory. For example, one readily verifies  that by averaging over the four configurations of Eq.~\eqref{eq:Configurations} all expectation values of the state  $|\downarrow\,\rangle$ are reproduced correctly.
        
\end{enumerate}
The number of equations that needs to be solved in this way scales only linearly with $N$ and therefore the dynamics of thousands of spins can be simulated. At the same time, the quantum mechanical uncertainties of the initial state are fully included by averaging over a distribution of initial spin vectors and in general $\langle \sigma_i^k \sigma_j^\ell \rangle\neq \langle \sigma_i^k\rangle\langle \sigma_j^\ell \rangle$. Importantly, since the mean-field equations of motion in Eq.~\eqref{eq:MF} preserve the length of each individual spin, $\vec s^{\,2}_i(t)=3$, its magnitude is equal to the exact quantum mechanical value along each individual trajectory. This means that compared to mean-field theory, the effects of spin-spin interactions, which scale as $\sim |\vec s_i(t)| |\vec s_j(t)|$, are much more accurately taken into account. In many situations of interest, for example, in spin-squeezing experiments with trapped ions or cold atoms, this feature leads to very precise predictions.  A more detailed discussion of the DTWA and many explicit examples can found in Refs.~\cite{Schachenmayer2015,Khasseh2020,Czischek2018,Zhu2019,Perlin2020}.

Finally, let us remark that instead of sampling from a discrete distribution, analogous simulations can be performed by sampling the initial spin components from a Gaussian distribution~\cite{Polkovnikov2010}, which leads to different levels of accuracies for different quantities and configurations~\cite{Sundar2019}. While we focus here exclusively on the DTWA, all results below can be also applied to such continuous TWA simulations.

\section{Stochastic simulation of open spin ensembles}\label{sec:stochastic}

In real experiments the spins or atoms are never completely isolated and will spontaneously decay or undergo dephasing due to residual interactions with the environment.  Such an open system scenario can be modeled by a master equation for the system density operator $\rho$~\cite{WallsMilburn,GardinerZoller}, 
\begin{equation}\label{eq:ME}
\dot{\rho}=-i [\mathcal{H},\rho]+ \mathcal{L}_{\rm deph}(\rho)+ \mathcal{L}_{\rm decay}(\rho).
\end{equation}
Here, the first correction to the Hamiltonian evolution accounts for pure dephasing, where for the case of uncorrelated dephasing of each spin with rate $\Gamma_\phi$ we obtain
\begin{equation}\label{eq:LindbladDeph}
\mathcal{L}_{\rm deph}(\rho)=\frac{\Gamma_{\phi}}{2} \sum_{i=1}^N \left(\sigma_i^z \rho \sigma_i^z - \rho\right).  
\end{equation}
In the other limit of interest, where the noise is fully correlated across the ensemble, we can use instead 
\begin{equation}
\mathcal{L}_{\rm deph}(\rho)= \Gamma_{\phi}^{\rm C} \left[2S_z \rho S_z - (S_z)^2\rho - \rho (S_z)^2 \right],
\end{equation}
where $S_z=\frac{1}{2}\sum_i \sigma^i_z$. The last term in Eq.~\eqref{eq:ME} is given by 
\begin{equation}
\mathcal{L}_{\rm decay}(\rho)= \frac{\Gamma}{2} \sum_{i=1}^N \left(2\sigma_i^- \rho\sigma_i^+ -  \sigma_i^+\sigma_i^- \rho - \rho  \sigma_i^+\sigma_i^- \right), 
\end{equation}
and describes the uncorrelated decay of each two-level system with rate $\Gamma$.

Naively, one could simply account for these decoherence processes by evaluating the mean-field dynamics for $\langle \sigma^k\rangle$ using the master equation in Eq.~\eqref{eq:ME} and by including the additional terms in the mean-field equations of motion in Eq.~\eqref{eq:MF}. This approach is still exact for noninteracting spins and would also in general correctly capture the decay of coherences of the transverse spin components, $\langle \sigma_i^x\rangle $ and $\langle \sigma_i^y\rangle$. However, in this case the spin length $|\vec s_i(t)|$ is no longer conserved along a trajectory. As a consequence, the effect of spin-spin interactions is also reduced and the accuracy of the DTWA simulation degrades considerably. 

In order to avoid this degradation, not only damping terms, but also an appropriate amount of fluctuations must be included in the dynamics. 
In a full quantum mechanical treatment, the added noise becomes most apparent when the open system dynamics is re-expressed in terms of quantum Langevin equations for the spin operators in the Heisenberg picture~\cite{WallsMilburn,GardinerZoller}. In previous works~\cite{Liu2020,Jaeger2021}, this representation has already been used to derive a set of semiclassical stochastic equations for the dynamics of (collectively) decaying two-level atoms. However, as explained in more detail in Appendix~\ref{app:QLE},  the mapping of quantum noise operators onto classical noise processes is ambiguous and in general leads to stochastic equations that are incompatible with DTWA simulations. Therefore, in the following we pursue two different approaches to identify suitable classical noise processes for modelling dephasing and decay in a spin-length preserving way.

\subsection{Dephasing} 
Let us first focus on pure dephasing. 
In this situation, instead of considering the Markovian master equation in Eq.~\eqref{eq:ME}, it is more convenient to start with the underlying dephasing interaction, which can be described by a  Hamiltonian of the form
\begin{equation}\label{eq:HamNoise}
\mathcal{H}_{\rm fluc}(t)= \frac{1}{2}\sum_{i=1}^N \xi_i(t) \sigma_i^z.
\end{equation}
Here the $\xi_i(t)$ are classical noise processes with zero mean and we can set $\langle \xi_i(t)\xi_j(t')\rangle\sim \delta_{ij} $ to model individual dephasing or $\xi_i(t)=\xi(t)$ for collective noise. The evolution under this Hamiltonian introduces an additional term in the mean-field dynamics,
\begin{equation}\label{eq:Stratonovich}
\left.\frac{d\vec s_i}{dt}\right|_{\rm deph}=  \xi_i(t) \vec e_z\times \vec s_i,
\end{equation}
i.e., a rotation around the $z$-axis with a fluctuating frequency.

\subsubsection{White noise limit}
Starting from this general noise model, we recover the Markovian dephasing dynamics in Eq.~\eqref{eq:ME} by considering the white-noise limit, where $\xi_i(t)$ is uncorrelated over the typical timescales of the spin dynamics. In this limit we can set
$\langle \xi_i(t)\xi_i(t')\rangle\simeq 2\Gamma_\phi\, \delta(t-t')$ and interpret Eq.~\eqref{eq:Stratonovich} as a Stratonovich stochastic differential equation. For numerical simulations it is more convenient to convert Eq.~\eqref{eq:Stratonovich} into an Ito differential equation, where the added noise in each time step is independent of $\vec s_i(t)$. Using the usual rules of stochastic calculus~\cite{Gardiner} we then obtain the following stochastic increments for the spin variables
\begin{eqnarray}
\label{eq:itodephasing1}
 \left.ds_i^x\right|_{\rm deph} &=& -\Gamma_{\phi} s_i^x dt- \sqrt{2 \Gamma_\phi} s_i^y dW_i,\\
 \left.ds_i^y\right|_{\rm deph} &=&-\Gamma_{\phi} s_i^y dt+ \sqrt{2 \Gamma_\phi} s_i^x  dW_i,\\
 \left.ds_i^z\right|_{\rm deph} &=& 0,
\label{eq:itodephasing3}
\end{eqnarray}
where the $dW_i\equiv dW_i(t)$ are real-valued and independent Wiener increments for the time step $[t,t+dt]$. These increments satisfy $\erw{dW_i}=0$ and $\erw{dW_i dW_j}=\delta_{ij} dt$ for individual dephasing and again we can simply set $dW_i=dW$ to describe spatially correlated noise. 

In summary, we end up with a DDTWA algorithm as illustrated in Fig.~\ref{fig:DDTWAscheme}(b). In this algorithm the sampling of the  initial spin values, $\vec s_i(t=0)$, is implemented as before, but the deterministic mean-field equations of motion for the dynamics are replaced by the following set of stochastic differential equations 
\begin{equation}
d\vec s_i = \vec \Omega_{\rm eff}^i \times \vec s_i  dt +  \left.d\vec s_i\right|_{\rm deph},
\label{eq:meanfielddeph}
\end{equation} 
where the dephasing-induced contribution is defined in Eqs.~\eqref{eq:itodephasing1}-\eqref{eq:itodephasing3}. This set of equations can be
efficiently simulated numerically with the Euler-Maruyama method~\cite{Gardiner}. We see that  Eq.~\eqref{eq:meanfielddeph} still describes the same coherent dynamics for $\langle \vec s_i\rangle(t)$, but also accounts for the loss of coherences.  Importantly, this loss is accompanied by an appropriate amount of noise, which ensures that~\cite{footnote}
\begin{equation}
\langle d \vec s^{\,2}_i \rangle = 0. 
\label{eq:spinlengths}
\end{equation}
Therefore, although coherences decay over time, the length of each spin and, as a consequence, also the magnitude of the spin-spin interactions are preserved on average.  In the examples discussed  in Sec.~\ref{sec:examples} below we find that this property results in an excellent agreement between these approximate stochastic simulations and the exact results obtained for a large variety of models and parameter regimes.

\subsubsection{Colored noise}
By modeling dephasing in terms of a fluctuating classical field, as in Eq.~\eqref{eq:HamNoise}, it also becomes straightforward to go beyond the Markov limit and generalize our results to colored noise without any approximations. To do so, we simply consider the evolution of the spins in the presence of noisy fields with a finite correlation time, for example,
\begin{equation}
\langle \xi_i(t)\xi_j(t')\rangle \simeq \delta_{ij} \sigma^2 e^{- \abs{t-t'}/\tau_{c}}.
\end{equation}
We see that in the limit $\tau_c \rightarrow 0$ we recover the $\delta$-correlated noise from above with $\Gamma_{\phi}=\sigma^2 \tau_{c}$, while for $\tau_c \rightarrow \infty$ we obtain the case of static noise with $\langle \xi_i(t)\xi_j(t')\rangle\simeq \delta_{ij}\sigma^2$.  In general, the random noises $\xi_i$ can be obtained by simulating an Ornstein-Uhlenbeck process~\cite{Uhlenbeck1930,Gardiner}
\begin{equation}\label{eq:ColoredIto}
d\xi_i=-\frac{1}{\tau_{c}} \xi_i dt+\sqrt{\frac{2}{\tau_{c}}} \sigma d\eta_i,
\end{equation}
where $d\eta_i$ are Wiener increments with $\erw{d\eta_i}=0$ and $\erw{d\eta_i d\eta_j}=\delta_{ij}dt$. In our numerical simulations we can then account for the effect of colored noise by simulating the coherent dynamics in Eq.~\eqref{eq:Stratonovich}, but assuming noisy fields $\xi_i(t)$ that are calculated according to Eq.~\eqref{eq:ColoredIto}. Note that compared to the Markovian case, this only increases the number of simulated equations by $N$ or even just by one in the case of collective noise. However, for very short correlation times $\tau_c$, also the integration time steps must be reduced and it becomes much more efficient to use the Markovian dephasing dynamics described by  Eqs.~\eqref{eq:itodephasing1}-\eqref{eq:itodephasing3}.

\subsection{Decay}
In the previous derivation we made use of the fact that dephasing can be described by classical noise. This is no longer the case for decay processes, where the system couples to a quantum environment represented by noise operators with non-vanishing commutation relations (see Appendix~\ref{app:QLE}). 
This difference between dephasing and decay processes also appears in the Schwinger-boson representation of collective spin systems, where in the latter case the mapping to a Fokker-Planck equation requires additional
approximations~\cite{Huber2021}.
In stochastic simulations of the full quantum mechanical wavefunction, decay is usually modelled by introducing random quantum jumps~\cite{GardinerZoller}, after which  the system is projected into the state of the spin pointing down, $|\downarrow\,\,\rangle$. Within the truncated Wigner function formalism, this would corresponds to a random projection into one of the four configurations listed in Eq.~\eqref{eq:Configurations}. However, in this approach the system evolution between the jumps is described by a non-Hermitian Hamiltonian, which again reduces the spin length $|\vec s_i|$ and may significantly degrade the accuracy of the DTWA (see Appendix \ref{app:Comparison}).

To circumvent these problems, we propose here to simulate the decay dynamics of dissipative spin systems by a continuous stochastic process with the following increments for the classical spin trajectories
\begin{equation}
d\vec s_i = \vec \Omega_{\rm eff}^i \times \vec s_i  dt + \left.d\vec
s_i\right|_{\rm decay},
\label{eq:meanfielddecay}
\end{equation}
where
\begin{eqnarray}
\label{eq:itodecay1}
\left.ds_i^x\right|_{\rm decay} &=& -\frac{\Gamma}{2} s_i^x dt-
\sqrt{\Gamma} s_i^y dW_i,\\
\left.ds_i^y\right|_{\rm decay} &=&-\frac{\Gamma}{2} s_i^y dt+ \sqrt{
	\Gamma} s_i^x  dW_i,\\
\left.ds_i^z\right|_{\rm decay} &=& -\Gamma (s_i^z+1) dt +
\sqrt{\Gamma} (s_i^z+1) dW_i.
\label{eq:itodecay3}
\end{eqnarray}
Let us emphasize that these equations are not derived from an underlying system-bath Hamiltonian, but rather constructed in order to satisfy two crucial properties. First, the deterministic terms in these equations reproduce the correct decay dynamics for the average spin components 
\begin{eqnarray}
\langle \dot  \sigma_i^{x,y}\rangle =&  -&\frac{\Gamma}{2} \erw{\sigma_i^{x,y}},\qquad \langle \dot  \sigma_i^z\rangle =  -\Gamma (\erw{\sigma_i^z}+1).
\end{eqnarray}
Second, the additional noise terms in Eqs.~\eqref{eq:itodecay1}-\eqref{eq:itodecay3} reintroduce spin fluctuations to preserve the length of each spin, $|\vec s_i|$, on average. 
 However, in contrast to the classical noise process, we now obtain~\cite{footnote} 
\begin{equation}
 \langle d \vec s^{\,2}_i \rangle = \Gamma \left(1-\erw{(s^z_i)^2}\right) dt,
\label{eq:spinlengthdecay}
\end{equation}
and this requirement can only be fulfilled up to a certain level of approximation. 
The reason is that for the decay process the deterministic change of the $z$-component, $d(s_i^z)^2=-2\Gamma s_i^z(s_i^z+1) dt$, is positive for $s_i^z<0$. This cannot be compensated by a positive diffusion term. In this sense, Eqs.~\eqref{eq:itodecay1}-\eqref{eq:itodecay3} represent a diffusion process, which reproduces the exact single-spin dynamics while conserving the length of each spin to a very good approximation.

In actual numerical simulations we find that in most situations of interest, in particular for small decay rates $\Gamma$, the condition $ \langle d \vec s^{\,2}_i \rangle\approx 0$ is satisfied and that the average spin length remains very close to its exact quantum mechanical value.  Specifically, in all the investigated examples reported in the following there are only  small variations in $\erw{\vec s_i^{\,2}}$ and therefore no noticeable degradation of the accuracy of the predicted results has been observed, neither in the transient dynamics nor in the steady state. While this condition cannot be guaranteed in general, the conservation of the spin lengths can easily be verified for a particular application. In this case, Eqs.~\eqref{eq:itodecay1}-\eqref{eq:itodecay3} represent a faithful stochastic approximation of a spin decay process, which is fully compatible with the DTWA. In addition, in Appendix~\ref{app:Improved} we present an alternative noise process based on two independent sets of Wiener increments. This stochastic update considerably improves the accuracy of the method for rates $\Gamma$ that are comparable to the timescales of the coherent evolution with only a modest increase in the computation time.

\section{Examples and benchmarking}\label{sec:examples}
In this section we demonstrate the application of the DDTWA for two paradigmatic settings in quantum optics, which can also be used to benchmark the results against exact numerical simulations in certain limiting cases. The first example is an ensemble of spin $S=1/2$ systems with spatially varying interactions. For this system it is already known that the DTWA provides accurate results in the isolated case~\cite{Schachenmayer2015} and we show that adding dephasing or decay does not affect the accuracy of the method. In the case of decaying spins we can also simulate the steady states of the ensemble and investigate, for example, non-equilibrium phase transitions in driven-dissipative spin systems. As a second setup we consider an ensemble of two-level atoms coupled to a common optical mode. This setting illustrates how the DDTWA can be combined with other phase space methods for continuous variable systems and shows that the relevant interplay between collective interactions and individual dephasing is accurately captured by these stochastic simulations.

\subsection{Interacting spin ensembles}
We first study the dynamics of an interacting spin ensemble under the influence of local dephasing and spontanenous emission, as described by Eqs.~\eqref{eq:ME}. More specifically, we assume that the  coherent evolution of the spins can be modeled by the transverse Ising Hamiltonian
\begin{equation}
\mathcal{H}= \frac{\Omega}{2} \sum_i \sx_i + \frac{1}{2}\sum_{i\neq j} J_{ij}  \sz_i \sz_j ,
\label{eq:HamZ}
\end{equation}
 where the spin-spin interactions,
\begin{equation}\label{eq:Jij}
J_{ij}=\frac{1}{N}\frac{J}{\abs{\vec r_i-\vec r_j}^{\alpha}},
\end{equation}
decay algebraically with the (normalized) distance between the spins, $\abs{\vec r_i-\vec r_j}$. Such a scenario appears, for example, in trapped ion systems, where $0<\alpha<3$~\cite{Bollinger2012,Richerme2014,Islam2013}, while for an ensemble of Rydberg atoms with van-der-Waals interactions we obtain $\alpha=6$~\cite{Labuhn2016,Bernien2017,Browaeys2020}.

\begin{figure}[t]
	\centering
	\includegraphics[width=1\columnwidth]{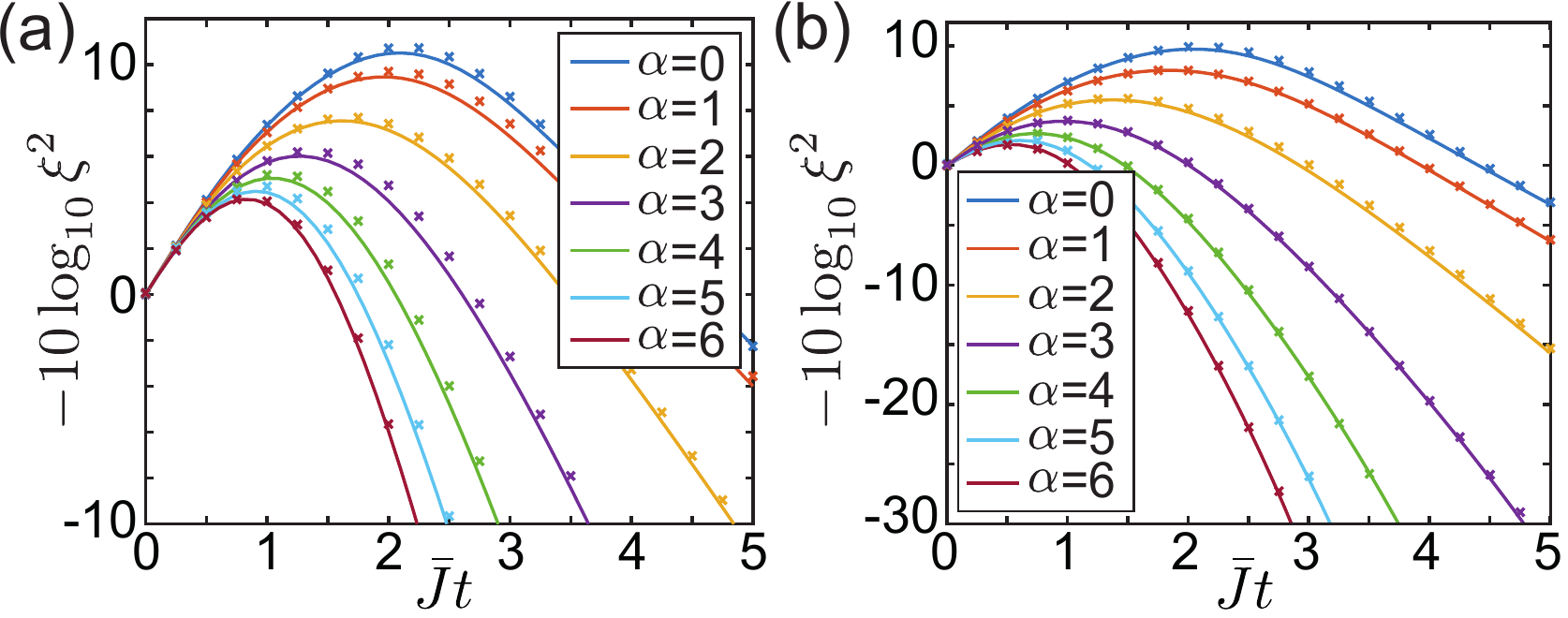}
	\caption{Time evolution of the squeezing parameter, $\xi^2$, for an ensemble of $N=64$ spins arranged on a 3D cubic lattice with unit spacing. The spins are initially aligned along the $x$-direction, $|\Psi_0\rangle=\prod_i |\rightarrow\,\rangle_i$, where $\ket{\rightarrow\,\,}=(\ket{\uparrow\,\,}+\ket{\downarrow\,\,})/\sqrt{2}$. For these simulations we assumed $\Omega=0$ and an individual dephasing of each spin with a rate (a) $\Gamma_\phi/J=0.0025$ and (b) $\Gamma_\phi/J=0.025$. For a better comparison the curves for different $\alpha$ are plotted in terms of the rescaled time unit $\bar J^{-1}$, where $\bar J=\sum_{i,j}J_{ij}/N$. The solid lines show the exact results~\cite{FossFeig2013} for different power-law interactions, as defined in Eq.~\eqref{eq:Jij}. The crosses show the corresponding values obtained with the DDTWA for $n_t=10000$ trajectories. }
	\label{fig:ZZDephasingAlpha}
\end{figure}

By adding the stochastic terms for local dephasing and spontaneous emission to the mean-field equations, we arrive at the following set of stochastic differential equations,
\begin{eqnarray}
ds_i^x&=&-\sum_{j\neq i} 2J_{ij} s_i^y s_j^z dt+ds_i^x|_{\rm deph}+ds_i^x|_{\rm decay},\\
ds_i^y&=& \sum_{j\neq i} 2J_{ij} s_i^x s_j^z  dt - \Omega s_i^z dt+ds_i^y|_{\rm deph}+ds_i^y|_{\rm decay},\\
ds_i^z&=& \Omega s_i^y dt+ds_i^z|_{\rm decay},
\end{eqnarray}
which we will now study in different limits of interest.

\subsubsection{Dephasing}

 \begin{figure}[t]
	\centering
	\includegraphics[width=1\columnwidth]{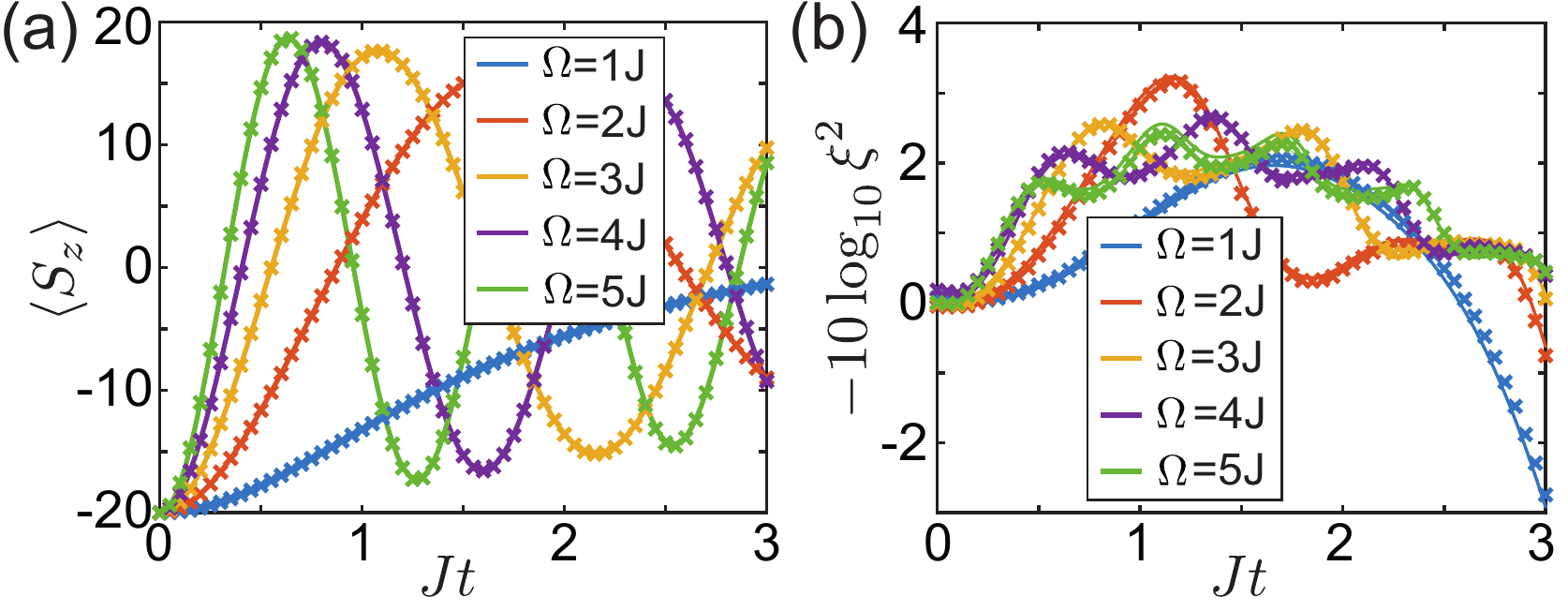}
	\caption{Plot of the time evolution of (a) the magnetization $\erw{S_z}$ and (b) the squeezing parameter $\xi^2$ of a driven spin ensemble with different driving strengths $\Omega$ and individual dephasing with rate $\Gamma_{\phi}/J=0.2$. Initially, all the spins are polarized along the negative $z$-axis, $|\Psi_0\rangle=\prod_i |\downarrow\,\rangle_i$. In both plots, $N=40$ and  all-to-all interactions ($\alpha=0$) are assumed. The solid lines are obtained from an exact integration of the master equation exploiting permutational invariance, while the crosses are obtained from a DDTWA simulation with $n_t=10000$ trajectories.}
	\label{fig:ZZDephasingOmega}
\end{figure}

In Fig.~\ref{fig:ZZDephasingAlpha} we use the DDTWA to evaluate, first of all, the dynamics of an interacting spin ensemble in the absence of the driving field, $\Omega=0$. In this example we have assumed that the $N=64$ spins are arranged on a cubic lattice in three dimensions with unit spacing and different values of the power-law exponent $\alpha$ are considered. The central quantity of interest in these plots is the spin-squeezing parameter~\cite{Wineland1992}
\begin{equation}
\xi^2=\min_{\phi} (\Delta S_{\phi}^{\perp})^2 \times \frac{N}{|\erw{\vec S}|^2}.
\end{equation}
Here $\vec S=(S_x,S_y,S_z)$ is the collective spin operator with components $S_k=\frac{1}{2} \sum \sigma_i^k$, and $S_\phi^\perp=\vec S\cdot \vec n_\phi^\perp$ is the projection of $\vec S$ onto an axis $\vec n_\phi^\perp$ parametrized by an angle $\phi$ in the plane orthogonal to the mean spin vector $\erw{\vec S}$. As usual, $(\Delta O)^2 = \erw{O^2}-\erw{O}^2$ denotes the variance of an operator $O$. Achieving a spin squeezing parameter of $\xi^2<1$ is relevant for metrological applications, but it also implies that the spins are entangled~\cite{Sorensen2001}. Therefore, such spin-squeezing effects cannot be described by mean-field theory.

In the absence of the driving field the $z$-components of all the spins are conserved and the system dynamics can still be evaluated exactly~\cite{FossFeig2013}. This allows us to directly compare the approximate stochastic simulations with the corresponding exact results. In Fig.~\ref{fig:ZZDephasingAlpha}(a) we find that for a very small dephasing rate of $\Gamma_{\phi}/J=0.0025$ the squeezing parameter $\xi^2$ calculated with the DDTWA is accurate up to the level of a few percent, which is consistent with DTWA results for isolated systems. As shown in Fig.~\ref{fig:ZZDephasingAlpha}(b), for a slightly stronger rate of $\Gamma_{\phi}/J=0.025$ the accuracy of the DDTWA improves even further. This can be attributed to the overall reduction of quantum correlations, which are only approximately take into account in the coherent dynamics.

In a next step we extend our analysis to finite driving strengths, $\Omega \neq 0$. In this case there are no analytic solutions available and exact numerical simulations are restricted to small spin systems, $N \lesssim 20$. However, in the limit of all-to-all interactions, i.e., $\alpha=0$, simulations with a large number of spins, $N \sim 100$, can still be done by exploiting the permutational symmetry of the master equation~\cite{Kirton2017,Shammah2018}. In Fig.~\ref{fig:ZZDephasingOmega} we use this symmetry to compare the DDTWA simulations of the driven Ising model with $\alpha=0$ to the corresponding exact numerical results. 
Again we find that for all the considered driving strengths the DDTWA provides very accurate predictions for the mean spin components as well as for the achievable level of quantum correlations signified by the squeezing parameter.

\subsubsection{Decay}
\begin{figure}[t]
	\centering
	\includegraphics[width=1\columnwidth]{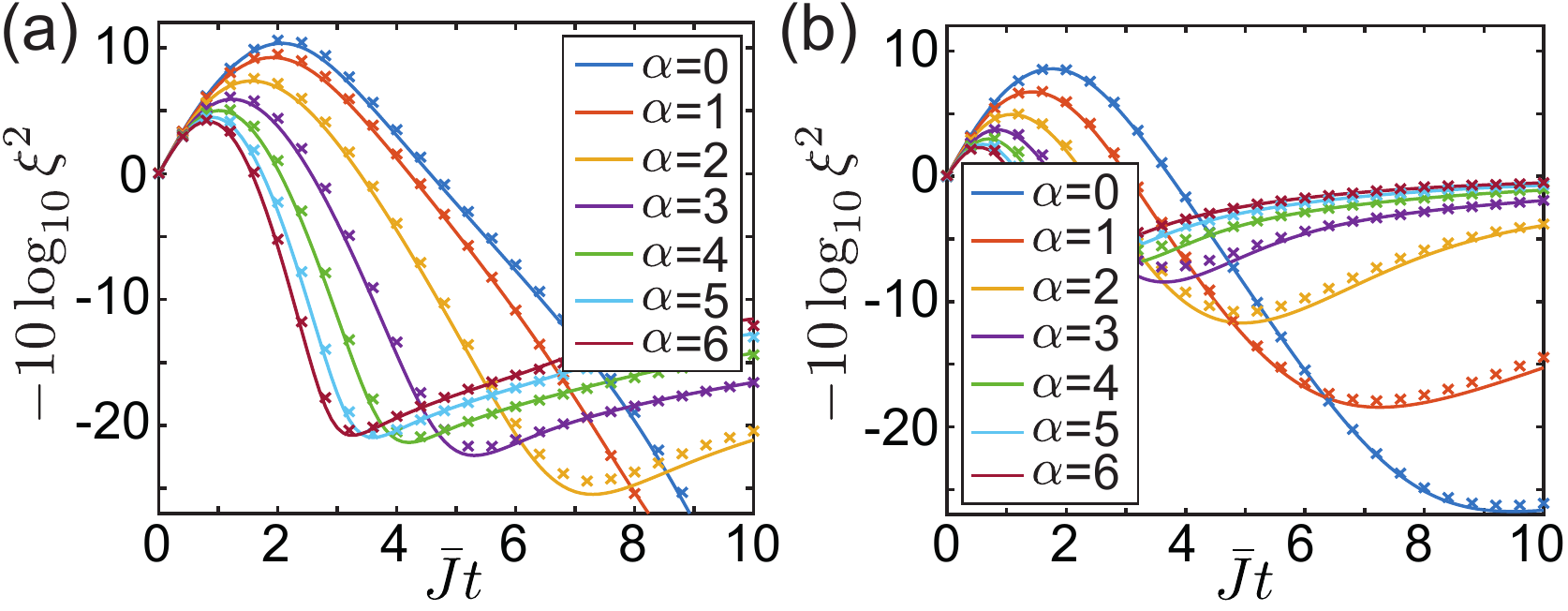}
	\caption{Time evolution of the squeezing parameter $\xi^2$ for an ensemble of $N=64$ spins arranged on a cubic lattice with unit spacing. The spins are initially aligned along the $x$-direction, $|\Psi_0\rangle=\prod_i |\rightarrow\,\rangle_i$, where $\ket{\rightarrow\,\,}=(\ket{\uparrow\,\,}+\ket{\downarrow\,\,})/\sqrt{2}$. For these simulations we assumed $\Omega=0$ and individual spontaneous emission of each spin with a rate (a) $\Gamma/J=0.0025$ and (b) $\Gamma/J=0.025$. 
		For a better comparison the curves for different $\alpha$ are plotted in terms of the rescaled time unit $\bar J^{-1}$, where $\bar J=\sum_{i,j}J_{ij}/N$. The solid lines show the exact results~\cite{FossFeig2013} for different power-law interactions, as defined in Eq.~\eqref{eq:Jij}. The crosses show the corresponding values obtained by the DDTWA method for $n_t=10000$ trajectories.}
	\label{fig:ZZEmissionAlpha}
\end{figure}
Let us now continue with a similar study of the transverse Ising model for $\Gamma_\phi=0$, but including a finite rate of decay, $\Gamma>0$. In Fig.~\ref{fig:ZZEmissionAlpha} we plot the spin-squeezing dynamics for different power-law interactions in the absence of the driving field, $\Omega=0$, and two different values of $\Gamma$. Similar to the case of dephasing, we find excellent agreement between the DDTWA simulations and the exact solutions~\cite{FossFeig2013}, which shows that for such short-time dynamical simulations both types of decoherence processes can be accurately taken into account. 

\begin{figure}[t]
	\centering
	\includegraphics[width=1\columnwidth]{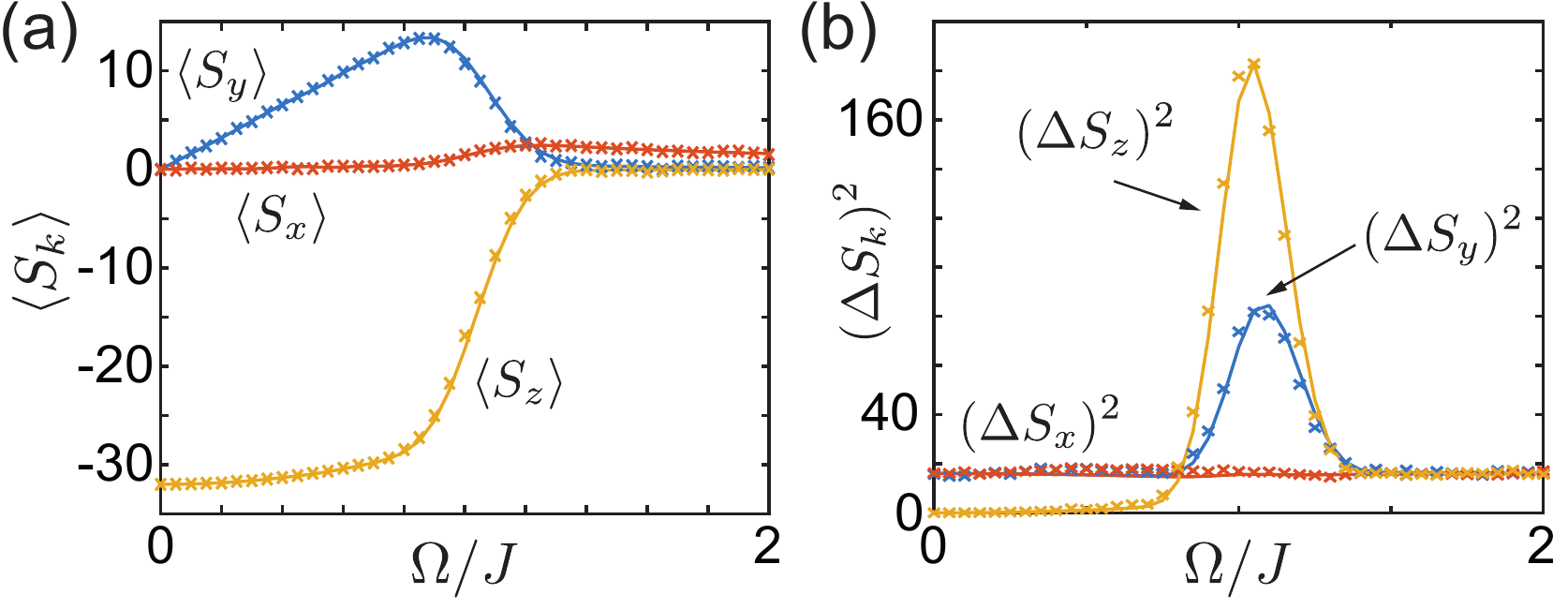}
	\caption{Steady state of the transverse Ising model given in Eq.~\eqref{eq:HamZ} with $\alpha=0$ and for a spin decay rate of $\Gamma/J=0.2$. The two plots show (a) the average values of the spin components, $\erw{S_{x,y,z}}$, and (b) their fluctuations, $(\Delta S_{k})^2=\langle S_k^2\rangle - \langle S_k\rangle^2$, as a function of the driving strength $\Omega$ and for $N=64$. The solid lines show the results from the exact simulation while the crosses were obtained from a DDTWA simulation by evolving $n_t=2000$ trajectories for a time $t=16/\Gamma$, starting from the initial state $|\Psi_0\rangle=\prod_i |\downarrow\,\rangle_i$.}
	\label{fig:TransverseIsingSSspontenousEmission}
\end{figure}

Let us now include a finite driving strength, $\Omega \neq 0$. While under the influence of pure dephasing the spin ensemble would then simply evolve into an infinite temperature state, this is not the case for driven spin systems in the presence of decay. As we illustrate in the following, the DDTWA can also be used to simulate such nontrivial steady states of driven spin ensembles. In order to benchmark these simulations, we focus again on the case $\alpha=0$, where exact numerical calculations are still possible. In Fig.~\ref{fig:TransverseIsingSSspontenousEmission} we evaluate the steady states of the dissipative transverse Ising model for varying driving strengths $\Omega$. For all parameters we  find excellent agreement between the DDTWA simulations and the exact results, both for the mean values of the collective observables $\erw{S_k}$ as well as for the variances $(\Delta S_k)^2$. The sharp peak in the spin fluctuations at a critical driving strength of $\Omega_c\approx J$ indicates a non-equilibrium phase transition in the steady state of the spin ensemble~\cite{Morrison2008}, which shows that the DDTWA is well suited to study such phenomena.  


In summary, these examples clearly demonstrate the high level of accuracy that can be achieved with the DDTWA when simulating interacting spin systems with local dephasing and decay and equivalently accurate results are obtained for spatially correlated dephasing. The remaining small deviations from the exact predictions do not vanish in the limit $\Gamma_\phi,\Gamma\rightarrow 0$ and can thus be attributed to inaccuracies in the coherent dynamics, where under the DTWA quantum correlations between the spins are only approximately taken into account. This is also consistent with the observation that in most situations the accuracy of the DDTWA  improves in the presence of decay or dephasing, where such quantum correlations are reduced.

\subsection{The driven Dicke model}\label{sec:Dicke}
Apart from being able to simulate large ensembles of spins, the DDTWA can be readily combined with conventional phase space methods for continuous degrees of freedom. This is relevant for a large range of cavity QED models, where many two-level systems are coupled to a common photonic mode. As an illustrative example, we consider here the driven Dicke model with Hamiltonian
\begin{equation}
\mathcal{H}=\frac{g}{\sqrt{N}}\left ( S_+ a + S_- a^\dagger \right) + \Omega S_x,
\label{eq:DickeH}
\end{equation}
where $S_\pm=S_x\pm i S_y$ and $a$ $(a^\dag)$ is a bosonic annihilation (creation) operator.  To model a realistic scenario, we include the dephasing of the two-level systems as well as the decay of  the photonic mode with a rate $2\kappa$. The whole system is then described by the master equation
\begin{equation}
\begin{split}
\dot{\rho}=-i [\mathcal{H},\rho]+ \mathcal{L}_{\rm deph}(\rho) + \kappa \left(2a\rho a^\dag - a^\dag a \rho - \rho a^\dag a\right).
\label{eq:DickeL}
\end{split}
\end{equation}
To apply the DDTWA in such a mixed setting, it is natural to represent also the bosonic mode in terms of its Wigner function,
\begin{equation}
W(\alpha, t)=\frac{1}{\pi^{2}}\int  d^{2} \beta \, e^{(\alpha \beta^*-\alpha^*\vec \beta )} \, {\rm Tr} \left\{ e^{\beta a^\dag-\beta^*a } \rho(t) \right\}.
\end{equation}	
In this case the moments of $W(\alpha, t)$ correspond to the symmetrically-ordered expectation values of mode operators~\cite{WallsMilburn,GardinerZoller}, 
\begin{equation}
\langle \{(a^\dag)^k a^\ell \}_{\rm sym}  \rangle (t) = \int d^n \alpha \,  (\alpha^{*})^k  \alpha^\ell W(\alpha,t).
\end{equation}
In the common situation where the photonic mode is initially prepared in the vacuum state or in a coherent state with amplitude $\alpha_0$, the corresponding Wigner function, 
\begin{equation}
W(\alpha, t=0) =\frac{2}{\pi} e^{-2\abs{\alpha-\alpha_0}^2}, 
\end{equation}
is positive and can be interpreted as a probability distribution for the classical amplitudes $\alpha$. In this case we can also sample the time evolution of $W(\alpha,t)$ by a set of stochastic trajectories   $\{\alpha_n(t)\}$ and evaluate expectation values as
\begin{equation}
\langle \{(a^\dag)^k a^\ell \}_{\rm sym}  \rangle (t) \simeq \frac{1}{n_t}\sum_{n=1}^{n_t} [\alpha_n^*(t)]^k  \alpha_n^\ell(t).
\end{equation}
In the absence of the two-level systems, these trajectories obey~\cite{GardinerZoller,WallsMilburn}
\begin{equation}
d\alpha |_{\rm loss}=-\kappa \alpha dt+\sqrt{\kappa/2} (dW_1 +i dW_2),
\end{equation}
and describe the loss of energy as well as the associated amount of quantum noise. 

Given a stochastic description for each of the individual subsystems, we can now simulate the dynamics of the whole setup by imposing a joint TWA, i.e., by treating the coupling between the photonic mode and the spins on a mean-field level. As a result we obtain the following set of stochastic differential equations
\begin{eqnarray}
ds_i^x&=& -\frac{2g}{\sqrt{N}} {\rm Im}(\alpha) s_i^z dt +ds_i^x|_{\rm deph},\\
ds_i^y&=&- \frac{2g}{\sqrt{N}} {\rm Re}(\alpha) s_i^z dt -\Omega s_i^z dt+ds_i^y|_{\rm deph},\\
ds_i^z&=& \frac{2g}{\sqrt{N}} \left[ {\rm Re}(\alpha) s_i^y+ {\rm Im}(\alpha) s_i^x \right]dt+ \Omega s_i^y dt,\\
d\alpha&=&-i  \frac{g}{\sqrt{4N}} \sum_i (s_i^x-i s_i^y) dt +d\alpha|_{\rm loss},
\end{eqnarray}
which are integrated for a set of  $n_t\gg1$ initial values $\vec s_i(0)$ and $\alpha(0)$, randomly drawn from the Wigner distributions of the individual subsystems.

\begin{figure}[t]
	\centering
	\includegraphics[width=1\columnwidth]{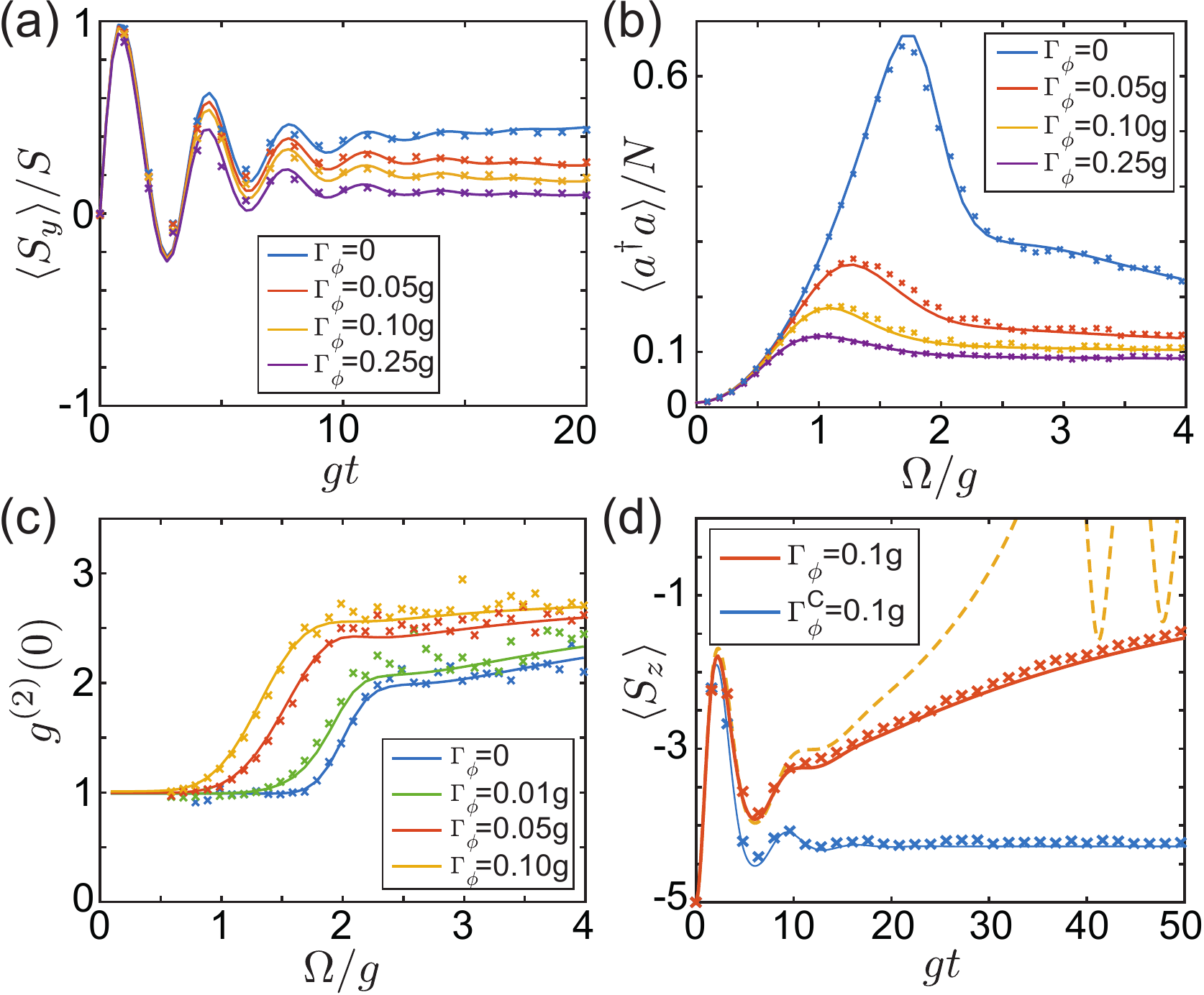}
	\caption{Simulation of the driven Dicke model as described in Eq.~\eqref{eq:DickeL} for $N=10$ and $\kappa=0.5g$, where the spins are initially prepared in the state $|\Psi_0\rangle=\prod_i |\downarrow\,\rangle_i$. 
		(a) Time evolution of $\erw{S_y}$ for $\Omega=2g$ and different dephasing rates $\Gamma_{\phi}$.
		(b) Steady state cavity occupation number $\langle a^\dag a\rangle$  and (c) steady state correlation function $g^{(2)}(0)$ as a function of the driving strength $\Omega$ and for different dephasing rates $\Gamma_{\phi}$.
		(d) Evolution of an initial fully polarized spin in the presence of individual ($\Gamma_\phi$) or collective ($\Gamma_\phi^{\rm C}$) dephasing with the same rate.  For this plot $\Omega=g$. In all the plots the solid lines represent the results obtained from an exact simulation of the master equation exploiting permutational invariance, while the crosses are obtained by the DDTWA method by simulating $n_t=2500$ trajectories. In (d) the dashed line shows the prediction from mean-field theory. }
	\label{fig:Dicke}
\end{figure}

In Fig.~\ref{fig:Dicke} we use this combined TWA approach to simulate the dynamics of the driven Dicke model, first of all for $N=10$ spins, where the results can still be compared with an exact simulation of the master equation. From this comparison we find an excellent agreement between the stochastic simulations and the exact results, both for the cavity and the spin observables. Although here we do not include a decay of the spins, the coupling to the lossy photonic mode still relaxes  the combined system [see Fig.~\ref{fig:Dicke}(a)]. Therefore, this setup also allows us to investigate the properties of the nontrivial steady states of this system. For example, in Fig.~\ref{fig:Dicke}(b) and (c) we plot the stationary value of the photon number and the two-photon correlation function 
\begin{equation}
g^{(2)}(0)=\frac{\erw{a^\dagger a^\dagger a a}}{\erw{a^\dagger a}^2}.
\end{equation}
In particular, this correlation function shows a qualitative change from a coherent state, where $g^{(2)}(0)\simeq 1$, to a thermal-like state with $g^{(2)}(0)\gtrsim 2$. This crossover as a function of the driving strength depends explicitly on the spin dephasing rate $\Gamma_\phi$. Note that in stochastic simulations, higher-order correlations typically have larger statistical errors, which can also be seen in the plot for $g^{(2)}(0)$.

As another illustrative example, we compare in Fig.~\ref{fig:Dicke}(d) the time evolution of the driven Dicke model for the two limiting cases of individual dephasing and collective dephasing with the same rate $\Gamma_\phi=\Gamma_\phi^{\rm C}$. For this plot we have assumed a moderate driving and coupling strength, such that the dissipative cavity acts mainly as a collective decay channel for the spins. For collective dephasing, where the system dynamics remains constrained to the maximal angular momentum subspace, the system then quickly relaxes to a stationary state with only a small spin population. In contrast, for individual dephasing the spin population increases with a rate $\sim \Gamma_\phi$ for longer times. This can be understood from the fact that the local dephasing processes drive the spins into orthogonal subspaces with a smaller total angular momentum quantum number. Within these subspaces there exist many  subradiant states, $|\psi_{\rm sub}\rangle$, which are decoupled from the cavity mode, i.e. $S_-|\psi_{\rm sub}\rangle=0$, but still have a finite spin population that remains trapped.

In Fig.~\ref{fig:Dicke}(d) we also show the prediction for $\langle S_z\rangle(t)$ obtained from mean-field theory. While mean-field theory still predicts very accurately the initial oscillations and the overall increase of the populations, the solution exhibits large, weakly-damped oscillations that are a clear artifact of this approximation. Note that the mean-field contribution to $\left.d\vec s_i\right|_{\rm deph}$ is the same for local and collective dephasing.  Thus, a mean-field simulation cannot distinguish between spatially correlated and uncorrelated noise, a difference that is manifested only in the stochastic noise terms.

In summary, the simulation of this driven Dicke model demonstrates the applicability of the DDTWA for simulating cavity QED systems with large ensembles of two-level systems. In particular, the example presented in Fig.~\ref{fig:Dicke}(d) shows that this method captures very accurately both the collective coupling to the maximal angular momentum states as well as the physics associated with subradiant states. We remark that beyond evaluating first- and second-order moments, simulations based on the TWA can in principle also be used to evaluate nonequal time correlation functions of the form $\langle O_A(t_2) O_B(t_1)\rangle$, following the general recipes discussed in Refs.~\cite{Polkovnikov2010,Berg2009,Wurtz2018,Deuar2021}. Although systematic benchmarks for such correlation functions are still required, the same ideas can be readily combined with the DDTWA to evaluate emission and absorption spectra~\cite{Liu2020,Jaeger2021} and other time nonlocal quantities.

\section{Large-scale simulations} \label{sec:largescale}
In the previous section we have focused on examples and parameter regimes where a comparison with other exact methods was still possible. However, the main advantage of the DDTWA is that it can be easily scaled up and applied in many experimentally relevant situations, where exact methods are no longer available. To illustrate this point, we consider in this section the superradiant decay of a large ensemble of interacting two-level atoms inside a lossy cavity.  An old question in connection to superradiance is how dipole-dipole interactions in dense atomic ensembles affect the decay process by inducing transitions out of the fully symmetric subspace~\cite{Gross1982}. In real experiments, similar effects can also arise from local dephasing and a relevant follow-up question is if interaction effects can actually be distinguished from fluctuating or static frequency inhomogeneities. As we show in the following, the DDTWA can be used to answer these and related  questions through direct numerical simulations.

 To do so we consider the same master equation as in Sec.~\ref{sec:Dicke},
\begin{equation}\label{eq:MELarge}
\begin{split}
\dot{\rho}=-i [\mathcal{H},\rho]+ \mathcal{L}_{\rm deph}(\rho) + \kappa \left(2a\rho a^\dag - a^\dag a \rho - \rho a^\dag a\right),
\end{split}
\end{equation}
but with a Hamiltonian of the form 
\begin{equation}\label{eqHLarge}
\mathcal{H}= \frac{g}{\sqrt{N}}\left ( S_+ a + S_- a^\dagger \right)  +\sum_{i<j} J^{xx}_{ij}  \sx_i \sx_j + \sum_i \frac{\omega_i}{2} \sigma_i^z.
\end{equation}
Here the first and the second term represent the collective atom-cavity coupling and the short-range spin-spin interactions with $J^{xx}_{ij}=J_x\abs{\vec r_i-\vec r_j}^{-3}$, respectively. The last term accounts for an inhomogeneous broadening of the atomic transition frequency, where the $\omega_i$ are randomly drawn from a Gaussian distribution with variance $\sigma^2$ and zero mean. 

\begin{figure}[t]
	\centering
	\includegraphics[width=1\columnwidth]{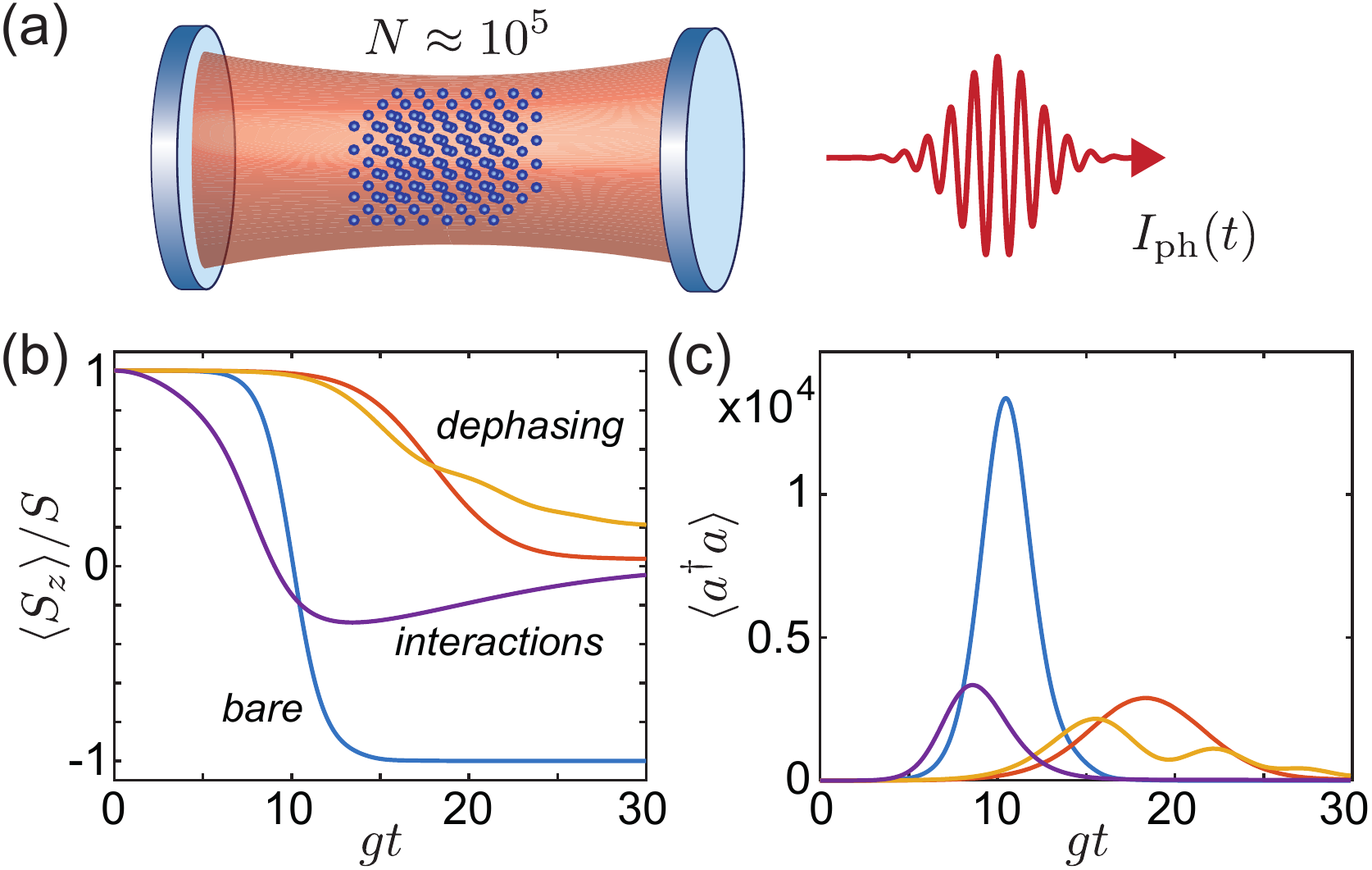}
	\caption{Superradiant decay of an ensemble of $N=47^3=103823$ two-level systems that are initially prepared in the excited state and couple to a lossy cavity mode with $\kappa=g$, as sketched in (a). In (b) we show the decay of z-magnetization and in (c) the photon number $\langle a^\dag a\rangle \sim I_{\rm ph}(t)$, which is proportional to the emitted field intensity. In both plots we compare the evolution of the bare non-interacting ensemble (blue line), with dynamics in the presence of additional Ising interactions $\sim \sigma_i^x\sigma^x_j$, with $J_x/g=0.025$ and $\alpha=3$ (purple line). The other two cases show the dynamics of a noninteracting ensemble, but in the presence of local Markovian dephasing with a rate $\Gamma_{\phi}/g=0.5$ (red line) and for an inhomogeneously broadened ensemble (yellow line).  In the latter case the spin frequencies $\omega_i$ have been randomly drawn from a normal distribution with zero mean and a variance of $\sigma^2=2\Gamma_\phi^2$. To obtain this data $n_t=64$ trajectories were simulated.}
	\label{fig:SuperradiantDecay}
\end{figure}


The model defined in Eq.~\eqref{eq:MELarge} and Eq.~\eqref{eqHLarge} can now be used to investigate, for example, how superradiant decay is influenced by (i) short-range interactions, (ii) Markovian dephasing and (iii) static inhomogeneous broadening. To do so we consider in Fig.~\ref{fig:SuperradiantDecay} a system of $N\approx 10^5$ atoms arranged on a cubic lattice and initially prepared in the excited state. We then use the DDTWA method to simulate the consecutive decay dynamics under the influence of those three processes. For these simulations we have assumed $\alpha=3$, but all interactions with $|J^{xx}_{ij}|/J_x<0.01$ have been set to zero. For the frequency distribution we have chosen a variance of $\sigma^2=2\Gamma_\phi^2$, such that the inhomogeneous broadening and the Markovian dephasing lead to a loss of coherence over a similar timescale. The plots in Fig.~\ref{fig:SuperradiantDecay} show that while all three mechanisms lead to a strong inhibition of the decay, the actual decay dynamics of the atomic population and the emitted photons is both qualitatively and quantitatively very different.

While a more detailed investigation of this system is beyond the scope of this work, these basic results already show how the DDTWA can be used to simulate interesting dynamical effects in large-scale spin systems under experimentally realistic conditions. Note that for the plots in Fig.~\ref{fig:SuperradiantDecay} we have simulated about $N=10^5$ atoms coupled to a cavity mode, which itself becomes populated with many thousands of photons. These simulations can be performed on a regular PC in about a day of computation time and with some additional programming efforts and the use of a supercomputer the simulation of millions of spins becomes possible.

Such system sizes are far beyond the typical atom numbers of about  $N\approx 150$~\cite{Kirton2017} that can be treated in exact simulations of similar models by exploiting permutation symmetry and using quantum trajectories. Moreover, both the short-range interactions as well as the inhomogeneous frequency distribution break the permutational invariance of the system such that the current type of simulations are simply not accessible with such exact numerical techniques. At the same time, since during the whole evolution $\langle \sigma^x_i\rangle=\langle \sigma^y_i\rangle=0$, neither the initial decay nor the effect of transverse spin-spin interactions would be captured by a simulation of the mean-field equations of motion only.  
Higher-order approximation schemes based on a cumulant expansion of correlation functions, which can account for such effects, already scale as $N^2$ and are thus no longer applicable for the considered system sizes. Note that cumulant expansion techniques also often exhibit numerical instabilities, which do not occur  in the DDTWA approach.


\section{Conclusion}\label{sec:conclusion}
In summary, we have presented a simple and efficient numerical algorithm for simulating large spin ensembles and cavity-QED systems in the presence of realistic decoherence processes. Using the DTWA for coherent systems as a starting point, we have shown that both dephasing and decay can be included in these simulations in terms of a stochastic evolution of the classical spin variables. Thereby it is possible to account for damping and loss of coherence while still preserving the total length of each classical spin on average. This last feature ensures that the magnitude of spin-spin interactions is not reduced and that the accuracy of the DTWA is not degraded. 

We have demonstrated and benchmarked the application of this method for various show cases, where a direct comparison with exact simulations is still possible. However, due to the linear scaling of the simulation time with the number of spins, the same results can be readily obtained for systems with many thousands or even millions of spins, which we have illustrated for the example of superradiant decay.  In such situations there are no exact numerical methods, but the DDTWA still allows us to make accurate predictions about fluctuations, correlations and spin-squeezing effects, as relevant for many cavity QED and spin ensemble experiments.  

\emph{Note added.} After the initial submission of this manuscript, a related work about combining the DTWA with the quantum jump formalism for open systems appeared~\cite{Singh2021}. A brief comparison between this method and DDTWA is given in Appendix~\ref{app:Comparison}. 

\acknowledgements
This work was supported through a DOC Fellowship (J.H.) from the Austrian Academy of Sciences (\"OAW) and by the Austrian Science Fund (FWF) through Grant No. P32299 (PHONED). A.M. acknowledges funding from NSF JILA-PFC PHY-1734006, NSF QLCI-2016244 and from the DOE-QSA grant.

\appendix

\section{Quantum Langevin Equations}\label{app:QLE}
In the main text we have introduced decoherence and dissipation effects via the master equation Eq.~\eqref{eq:ME} for the system density operator. In general, for a master equation of the form
\begin{equation}
\dot \rho= -i[\mathcal{H},\rho] +\frac{\gamma}{2}\left(2C\rho C^\dag - C^\dag C \rho - \rho C^\dag C\right),
\end{equation}
with jump operator $C$, there exists an equivalent quantum Langevin equation~\cite{GardinerZoller},
\begin{equation}\label{eq:QLE}
\begin{split}
\dot O= i[\mathcal{H},O] &- [O,C^\dag] \left(\frac{\gamma}{2}C+\sqrt{\gamma} \hat f_{\rm in}(t)\right) \\
&- \left(\frac{\gamma}{2}C^\dag+\sqrt{\gamma} \hat f^\dag_{\rm in}(t)\right)[C,O],
\end{split}
\end{equation}
which describes the corresponding dynamics of a system operator $O(t)$ in the Heisenberg picture. In this equation, the noise operators $\hat f_{\rm in}(t)$ and $\hat f^\dag_{\rm in}(t)$ satisfy $[\hat f_{\rm in}(t),\hat f^\dag_{\rm in}(t')]=\delta (t-t')$ and~\cite{GardinerZoller} 
\begin{equation}\label{eq:NoiseCommutator}
[O(t),\hat f_{\rm in}(t) ]= \frac{\sqrt{\gamma}}{2} [ C(t),O(t)].
\end{equation}
This last relation implies that the noise and the system operators are not independent of each other, similar to a Stratonovich stochastic differential equation. Consistent with this interpretation, the quantum Langevin equations obey the rules of regular calculus, such as the product rule.

Given the original derivation of the DTWA in terms of a factorization approximation for Heisenberg operators~\cite{Schachenmayer2015}, it is tempting to use the quantum Langevin equations for the spin operators $O=\{\sigma_i^k\}$ as a starting point for deriving the stochastic equations of motion for the DDTWA in open systems. However, due to their nontrivial commutation relations there is a priori no obvious procedure how to replace the quantum noise operator $\hat f_{\rm in}(t)$ by a suitable classical noise process. 

To illustrate this point, let us first consider the evolution of a single spin under pure dephasing, where  $\gamma=\Gamma_\phi/2$, $C=\sigma^z$ and 
\begin{eqnarray}
\dot \sigma^x &=&- \Gamma_\phi \sigma^x + i \sqrt{2\Gamma_\phi} \left( \sigma^y \hat f_{\rm in}(t) - \hat f_{\rm in}^\dag(t) \sigma^y\right),  \\
\dot \sigma^y &=& - \Gamma_\phi \sigma^y - i \sqrt{2\Gamma_\phi} \left( \sigma^x \hat f_{\rm in}(t) - \hat f_{\rm in}^\dag(t) \sigma^x\right),  \\
\dot \sigma^z&=&0. 
\end{eqnarray}
We emphasize that under the validity of the Markov approximation, these equations are still exact. However, in order to derive semiclassical equations of motion compatible with DTWA simulations, we must replace the operator $\hat f_{\rm in}(t)$ by classical noise. In the case of dephasing, the naive substitution $\hat f_{\rm in}(t)\rightarrow \xi(t)=[\xi_1(t)+i\xi_2(t)]/\sqrt{2}$, where $\langle \xi_i(t) \xi_j(t')\rangle=\delta_{ij} \delta(t-t')$ are real-valued classical noise processes, would already result in equations similar to Eqs.~\eqref{eq:itodephasing1}-\eqref{eq:itodephasing3}, but with a diffusion constant that is off by a factor of two. This factor can be corrected  by using  the substitution $\hat f_{\rm in}(t)\rightarrow \xi(t)/\sqrt{2}$ instead. Note, however, that in order to arrive at the final stochastic equations in Eqs.~\eqref{eq:itodephasing1}-\eqref{eq:itodephasing3}, it also requires an additional ad hoc reinterpretation of the original Stratonovich equations as Ito stochastic equations with independent noise increments. Therefore, even for this simple example, the mapping between quantum and classical noise introduces various ambiguities that basically require the knowledge of the correct result in advance.

For the evolution of a single spin subject to decay we identify  $\gamma=\Gamma$ and $C=\sigma^-$. In this case we obtain 
\begin{eqnarray}
\dot \sigma^x &=& - \frac{\Gamma}{2} \sigma^x +\sqrt{\Gamma}\left( \sigma^z \hat f_{\rm in}(t) + \hat f_{\rm in}^\dag(t) \sigma^z\right),  \\
\dot \sigma^y &=&  - \frac{\Gamma}{2} \sigma^y  +i \sqrt{\Gamma}\left( \sigma^z \hat f_{\rm in}(t) - \hat f_{\rm in}^\dag(t) \sigma^z\right),\\
\dot \sigma^z&=&-\Gamma(\sigma^z+\mathbbm{1}) - 2\sqrt{\Gamma}\left( \sigma^+\hat f_{\rm in}(t) + \hat f_{\rm in}^\dag(t) \sigma^-\right).
\end{eqnarray}
We can follow now the same procedure as above and after substituting $\hat f_{\rm in}(t)\rightarrow \xi(t)/\sqrt{2}$, taking expectation values for the spin variables and interpreting the resulting equations as Ito differential equations we arrive at the following set of classical equations 
\begin{eqnarray}
\label{eq:itodecayapp1}
d s^x &=& - \frac{\Gamma}{2} s^x dt  +\sqrt{\Gamma} s^z dW_1(t), \\
d s^y &=&  - \frac{\Gamma}{2} s^y dt - \sqrt{\Gamma} s^z dW_2(t), \\
d s^z&=&-\Gamma(s^z+1) dt - \sqrt{\Gamma} \left[s^x dW_1(t) - s^y dW_2(t)\right].
\label{eq:itodecayapp3}
\end{eqnarray}
We see that the noise terms in these equations differ qualitatively from the ones in Eqs.~\eqref{eq:itodecay1}-\eqref{eq:itodecay3}. Importantly, according to Eqs.~\eqref{eq:itodecayapp1}-\eqref{eq:itodecayapp3} the average spin length changes in each time step as 
\begin{equation}
 \langle d \vec s^{\,2} \rangle = - 2\Gamma \erw{s^z} dt,
\end{equation}
and is thus not conserved. This means that over time, spin fluctuations will considerably deviate from their exact quantum mechanical value and a faithful simulation of correlations or of steady states becomes impossible (see the following discussion in Appendix~\ref{app:Comparison}).

In conclusion, these examples illustrate that there is no simple procedure to map quantum noise operators onto equivalent classical noise processes that are suited for DDTWA simulations. 
In the case of dephasing we have  circumvented this problem by considering from the start a classical noise model, see Eq.~\eqref{eq:HamNoise}. For the case of decay, we have identified an approximate noise process guided by the key requirement of conserving the fluctuations of each individual spin. However, this choice is not unique and in Appendix~\ref{app:Improved} we present an alternative stochastic processes, which further improves the accuracy of the simulation.

%
%

\section{Importance of length-preserving noise processes in long-time simulations} \label{app:Comparison} 

\begin{figure}[t]
	\centering
	\includegraphics[width=\columnwidth]{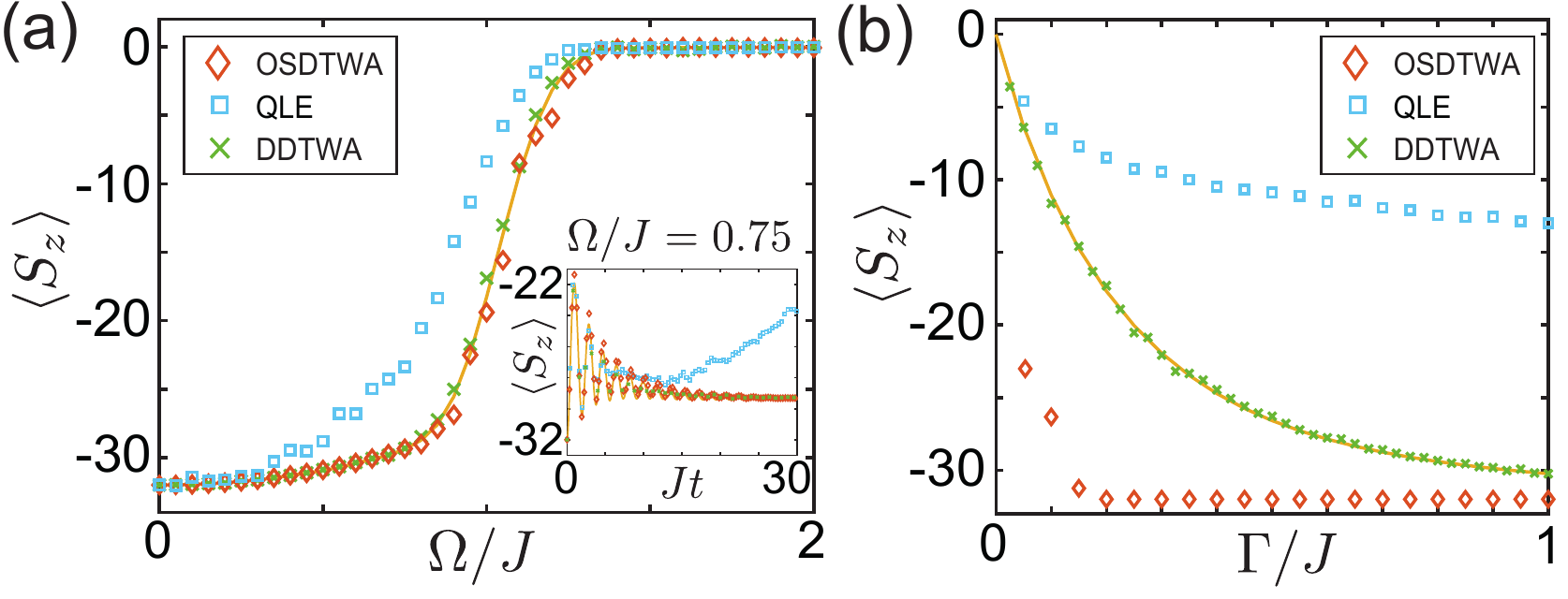}
	\caption{Comparison of DDTWA with other semiclassical simulation schemes, QLE and OSDTWA, where the magnitude of each classical spin is not conserved on average. (a) Steady state of the transverse Ising model with the same parameters as in Fig.~\ref{fig:TransverseIsingSSspontenousEmission}. The inset shows that the QLE method does not reach a steady state. (b) Magnetization $\erw{S_z}$ at time $t=40/J$ for a spin model with Hamiltonian $H= J  S_x^2/N$ and individual spin decay $\Gamma$. In both plots $N=64$ and the solid lines represent the exact results. }
	\label{fig:comparision}
\end{figure}

In Ref.~\cite{Liu2020} a derivation similar to the one outlined in Appendix~\ref{app:QLE} has been used to derive noise processes for collective and individual spin decay. In the lasing simulations described in this work, the relevant interaction time between each two-level atom and the optical mode is short compared to $\Gamma^{-1}$ such that the influence of the added noise is negligible. However, the noise terms proposed in Ref.~\cite{Liu2020} either do not conserve the length of the individual spins or result in negative diffusion terms. Thus they are not suited for simulating the long-time dynamics or steady states of open spin systems. Related problems can occur in simulations based on quantum jumps \cite{Singh2021}, where during the intermediate periods of non-unitary evolution, the spin length decays. 

To illustrate this important issue, we compare in Fig.~\ref{fig:comparision}  the DDTWA method with (i) semiclassical simulations of the quantum Langevin equations (QLE), where decay is described by the noise processes in Eqs.~\eqref{eq:itodecayapp1}-\eqref{eq:itodecayapp3}, and (b) with the quantum jump method OSDTWA described in Ref.~\cite{Singh2021}. In Fig.~\ref{fig:comparision} (a) we first reproduce the results of the dissipative transverse Ising model already shown in Fig.~\ref{fig:TransverseIsingSSspontenousEmission}. In Fig.~\ref{fig:comparision} (b) we consider another, even simpler model with Hamiltonian $H= J  S_x^2/N$ and individual spin decay.  In both examples we find a strong discrepancy between the QLE approach and the exact results, and the simulations don't reach a steady state. This can be attributed to constantly growing spin fluctuations, $\langle \vec s_i^{\,2}\rangle(t) \gg 3$, which leads to a strong overestimation of interaction effects. The OSDTWA method still performs very well in the first example, but it completely misses the effect of quantum fluctuations in the second model. In this case the final state approaches $\langle S_z\rangle=-N/2$ for all values of $\Gamma/J$.   In this case one finds $\langle \vec s_i^{\,2}\rangle(t) \ll 3$. In both examples, the DDTWA, where $\langle \vec s_i^{\,2}\rangle(t) \approx 3$, provides accurate results.

\section{Improved noise process for simulating decay}\label{app:Improved}
\begin{figure}[t]
		\centering
		\includegraphics[width=1\columnwidth]{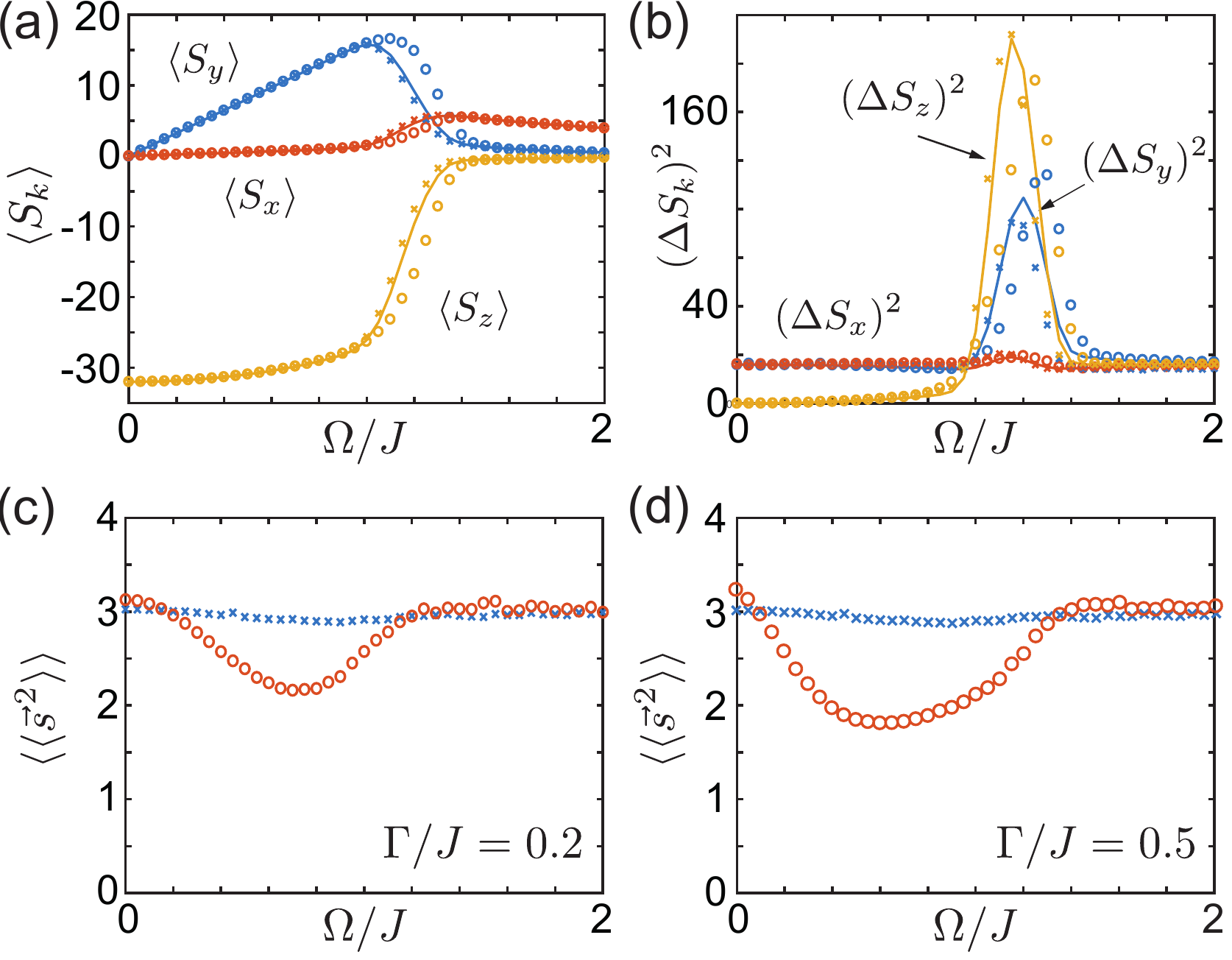}
		\caption{(a) and (b) The same plots as in Fig.~\ref{fig:TransverseIsingSSspontenousEmission} (a) and (b), but for a larger value of $\Gamma/J=0.5$. (c) and (d) Variation of the corresponding average spin length in steady state, $\langle\langle \vec s^{\,2}\rangle\rangle=\sum_i \langle \vec s_i^{\,2}\rangle/N $, for two different values of $\Gamma$. In all plots the solid lines represent  the exact results and the circles mark the DDTWA simulations based on the stochastic updates given in Eqs.~\eqref{eq:itodecay1}-\eqref{eq:itodecay3}. The crosses indicate the results obtained via the improved noise process defined in Eqs.~\eqref{eq:itodecay1n}-\eqref{eq:itodecay3n}. All other parameters are the same as in Fig.~\ref{fig:TransverseIsingSSspontenousEmission}.}
		\label{fig:TransverseIsingSSspontenousEmissionComparison}
\end{figure}

In Eqs.~\eqref{eq:itodecay1}-\eqref{eq:itodecay3} we have identified  a set of stochastic equations for simulating the independent decay of each spin. As emphasized in the main text, these equations are only approximate and in many cases the accuracy of the simulations degrades when $\Gamma^{-1}$ becomes comparable to the timescale of the coherent evolution. For example, in Fig.~\ref{fig:TransverseIsingSSspontenousEmissionComparison} (a) and (b) we re-perform the same steady state simulation as in Fig.~\ref{fig:TransverseIsingSSspontenousEmission},  but with a higher decay rate of $\Gamma/J=0.5$. In this case we see clear quantitative deviations between the stochastic simulations and the exact results. In Fig.~\ref{fig:TransverseIsingSSspontenousEmissionComparison} (c) and (d) we also plot the steady-state expectation value of the average spin length, $\langle\langle \vec s^{\,2}\rangle\rangle=\sum_i \langle \vec s_i^{\,2}\rangle/N $, as obtained from the DDTWA simulations for two values of $\Gamma/J$. While for very weak and strong driving the spin length is almost conserved, $\langle\langle \vec s^{\,2}\rangle\rangle\simeq 3$, there are significant deviations from this value for $\Gamma \approx \Omega$.

In order to improved the accuracy of the DDTWA also for intermediate decay rates, we present here an extended noise process for simulating decay, which for the $i$-th spin reads:
\begin{eqnarray}
\label{eq:itodecay1n}
ds_i^x&=& -\frac{\Gamma}{2} s_i^x dt -\frac{\sqrt{\Gamma}\left[(s_i^y+1) dW_i^1+(s_i^y-1) dW_i^2\right]}{2},\\
ds_i^y&=&-\frac{\Gamma}{2} s_i^y dt+\frac{\sqrt{\Gamma}\left[(s_i^x+1) dW_i^1+(s_i^x-1) dW_i^2\right]}{2},\\
ds_i^z&=& -\Gamma (s_i^z+1) dt +
\sqrt{\frac{\Gamma}{2}} (s_i^z+1) (dW_i^1-dW_i^2).
\label{eq:itodecay3n}
\end{eqnarray}
Here $dW_i^1$ and $dW_i^2$ are two independent Wiener increments. According to these equations, the change of the average spin length is now given by 
\begin{equation}
\langle d \vec s^{\,2}_i \rangle = \frac{\Gamma}{2} \left[4-\erw{(s^x_i)^2}-\erw{(s^y_i)^2}-2\erw{(s^z_i)^2}\right] dt,
\end{equation}
where, compared to Eq.~\eqref{eq:spinlengthdecay}, the three spin projections appear more symmetrically.  As shown by the simulation results indicated  by the crosses in Fig.~\ref{fig:TransverseIsingSSspontenousEmissionComparison}, this fact leads to a considerable improvement of the accuracy of the steady-state mean values and variances. We also see that the average spin length is now fully conserved for all values of $\Gamma$. Note, however, that this comes at the prize of doubling the amount of random numbers that are required at each time step and for many applications of interest, where $\Gamma$ is anyway small, the simpler noise increments in Eqs.~\eqref{eq:itodecay1}-\eqref{eq:itodecay3} are still sufficient.

\begin{thebibliography}{99}
\bibitem{footnote} For a stochastic variable $x$ with increment $dx =- a dt + b dW$, one can write  $dx^2=(x+dx)^2-x^2$ and apply  the Ito rules $\langle dW\rangle =0$ and $\langle dW^2\rangle=dt$ in order to obtain $\langle dx^2\rangle =(- 2a x + b^2) dt$.
\end{thebibliography}
\end{document}